# The yielding of concentrated cohesive suspensions can be deformation rate-dependent.

Richard Buscall[1,2], Peter J Scales[2], Anthony D Stickland[2], Hui-En Teo[2], Tiara E Kusuma[2], Sayuri Rubasingha[2] & Daniel R Lester.[3,4.]

[1]MSACT Research and Consulting, Exeter, EX2 8GP, UK.
[2]Dept. Chemical and Biomolecular. Engineering, University of Melbourne, Australia 3010.
[3]CSIRO Mathematics, Informatics and Statistics, Highett, Australia 3190.
[4]School of Civil, Environmental & Chemical Engineering, RMIT, Melbourne, Australia 3001.

Empirical models of yield stress liquids in steady shear-flow such as the popular Herschel-Bulkley model partition the total stress into two parts: a part associated with the solid phase, the yield stress, and a viscous part, which, in Herschel-Bulkley [1], is power-law shear-thinning in general. In simple shear then, the Herschel-Bulkley equation can be written as $\sigma = \sigma_0 + k\dot{\gamma}^n$, the simpler Bingham model being recovered in the special case of $n$ =1. The yield stress or solid phase part $\sigma_0$ accounts for stress transmission by direct inter-particle interactions, notionally, whereas the viscous part accounts for dissipation in the suspending liquid and the effect of the particles upon it. Viscous shear thinning ($n$ < 1) can be interpreted in terms of a decrease of effective volume-fraction with increasing shear-rate.

My no means all cohesive suspensions show simple Herschel-Bulkley behaviour. The apparent yield stress can depend upon how one sets out to measure it [2,3,5]. It can be very irreproducible too, even within the scope of a single test protocol [3]. Furthermore, some cohesive suspensions display highly non-monotonic flow curves [3,4]. We have recently found a suspension that will show any of the aforementioned behaviours, depending upon the type of test protocol used. The material is a concentrated suspension of 4.5 μm $CaCO_3$ particles in water, coagulated at the IEP of $CaCO_3$ [3,5,8].

The table below lists some of the test modes used and the behaviours associated with them. Note that 'CR' therein denotes 'controlled rate', whereas 'CS' mean 'controlled stress'. $Pe_0$ is the so-called "bare" Péclet number $6\pi R^3\mu\dot{\gamma}/k_BT$ [9], where $R$ is the mean particle radius, $\mu$ the viscosity of the liquid phase, $T$ is absolute temperature and $k_B$ is Boltzmann's constant.

**Table 1: Yield behaviour depends upon test type.**

| | Test protocol | | Behaviour |
|---|---|---|---|
| **A** | An ascending "staircase" of rates in time, all at $Pe_0 > 1$. | CR | Herschel-Bulkley [3]. |
| **B** | As above but starting from $Pe_0 << 1$ | CR | Non-monotonic flow curve [3]. |
| **C** | Creep testing at a series of stresses. | CS | Time-dependent yield over a modest range of stress [5]. |
| **D** | An ascending "staircase" of stresses in time (CS flow curve). | CS | Erratic yield and shear banding [3,5]. |
| **E** | As above but with a return down the staircase of stresses. | CS | Hysteresis between ascending and descending branches [3]. |

We found that the response changed from test to test because the yield stress was deformation-rate dependent. It was so to the point where, at a volume-fraction of 0.40 for example, the apparent yield stress varied from < 5 Pa to ca. 200 Pa, depending upon the method used and the rate of deformation associated with it [3,5]. Appreciable variation from one method to another has been reported before (e.g. [2]) but nothing quite on this scale perhaps. The yield *strain* varied with deformation rate too, from a value of ca. unity at low $Pe_0$, a value to be associated with cooperative local or "cage" melting according to Pham et al. [2] and others, down to a value of ca. $10^{-4}$ at $Pe_0 \sim 1$ and above, this being the magnitude of strain needed to break interparticle bonds in the $CaCO_3$ suspension [8].

The diversity of behaviour summarised in table 1 could be rationalised in terms of deformation-rate induced melting of the local structure [8]. Furthermore, for steady-state purposes, all that one needs to do in order to account for it is to modify the Herschel-Bulkley equation by incorporating a second shear-rate or Pe-dependent function, together with a new constant thus,

$$\sigma = \sigma_0 g(Pe_0) + \sigma_{iso} + k\dot{\gamma}^n \,, \qquad (1$$

where, in simple shear-flow, the function $g(Pe_0)$ is a decreasing function of shear rate; one that decays to zero at some point, and where the constant $\sigma_{iso}$ is introduced to recover Herschel-Bulkley behaviour as a limiting case at high $Pe_0$.

Flow curves for the $CaCO_3$ suspension at a volume-fraction of 0.40 are shown in Fig. 1. The figure displays data obtained from tests of types B, D and E above. The data were obtained using cruciform cross-section vanes immersed in wide cylinders giving wide gaps [3,5]. The stress is plotted against the logarithm of the angular velocity, the *apparent*, or Newtonian-equivalent shear-rate at the vane being a little over twice this.

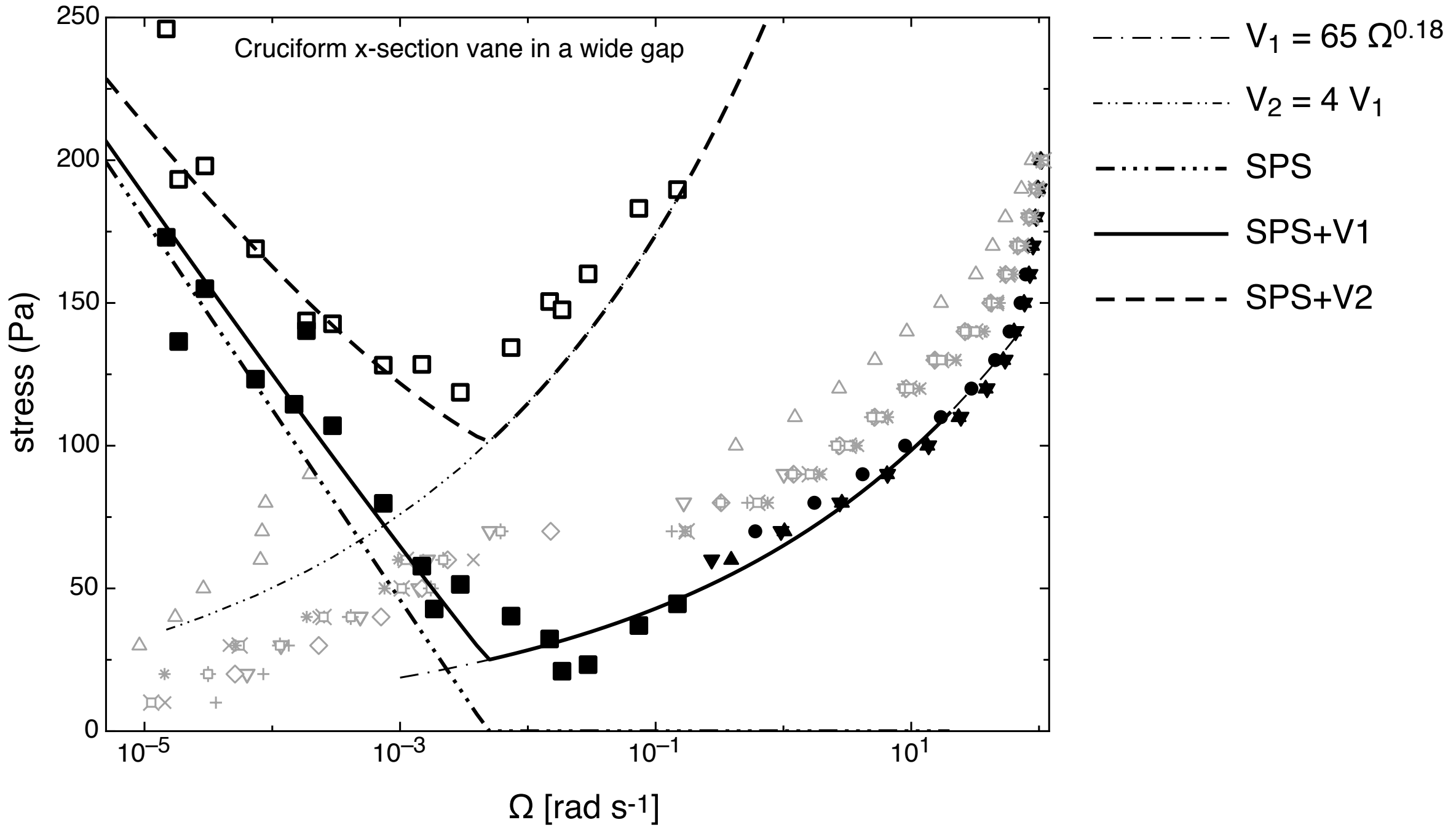

**Fig. 1** Composite flow curve for a 40%v/v coagulated $CaCO_3$ suspension. The smaller black points are from CS in descent (cf. "E" in the table above). The greyed out symbols are the erratic CS ascent data (cf. "D"). The larger filled and unfilled black squares are the CR steady-state and peak stresses respectively (cf. "B"). The flow curves have been fitted as shown (cf. eqn 2). The apparent or Newtonian equivalent shear rate is approximately twice the rotation rate. On the ascending branch the true shear rate is ca. 5.6 times that. Data taken from reference 3.

The calculation of the true shear-rates on the left-hand, descending branch is problematic even in the case of controlled-rate where there can be no shear banding. The right-hand, or viscous branch is however straightforward as here all one needs to do is to divide the apparent shear-rate by the power-law index, *n*. On the left-hand branch the material behaves like a different Herschel-Bulkley liquid at each point in effect; one with a different yield stress at each shear-rate. The way that we have solved the problem of estimating the true shear rate is to suppose that the viscous power-law fit to the right-hand branch can be continued leftwards, i.e., that the total stress on the left hand branch is given by the sum of a solid-phase contribution, 'SPS' in the legend, and a power-law viscous contribution, '$V_1$' as shown. The vane shear-rates can then be calculated. The results are model-dependent, of course, except that there is no avoiding this there being no model independent way of attacking the problem.

The dependence of the solid phase stress on the apparent and corrected shear-rates is shown in fig. 2. The curves imply that the true strain-rate softening function $g(\dot{\gamma})$ producing the flow curve could well be much stronger than it first appears. One needs to be cautious about this conclusion though: stress growth measurements in step shear-rate [8] suggested that the solid-phase stress reached a constant value at an apparent Hencky strain (or, equivalently, scaled time for stress growth in step strain rate) of $\ln(1+\dot{\gamma}t) \sim 0.5$ and it simply cannot be assumed that the flow is fully developed at such strains. Indeed, it was found that it was not, since the viscous stress peaked there before decaying to a somewhat noisy steady-state value at a Hencky strain of ~ 3 [8]. The underlying or true dependence of the solid phase stress on shear rate is thus thought to be closer to the continuous line. Evidence for this will be presented in the next article when we will discuss the transient behaviour in more detail.

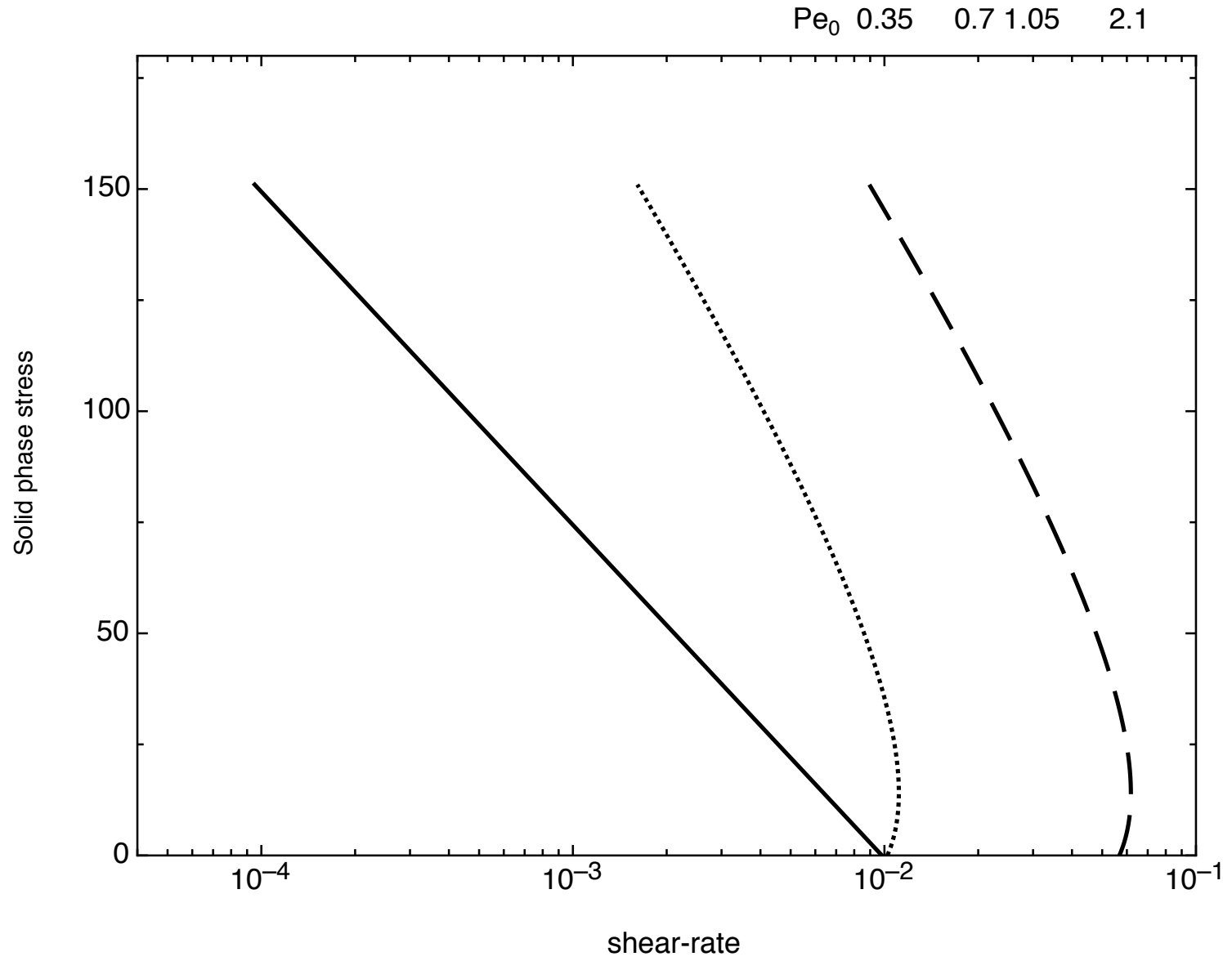

**Fig. 2** Solid-phase stress (SPS from fig.1) plotted against shear-rate. Continuous line – apparent, Newtonian, shear-rate, dashed line – against the corrected steady-state shear-rate. The dotted line merely serves to show that part of the difference between the former two curves comes from the power law, i.e. a 1/*n* shift, the remaining difference is due to the fact that the yield stress is rate-dependent. The true dependence of SPS on shear rate is believed to be more like the continuous line because it plateaus at strain ~0.5 where the flow is not fully developed.

The peak stress (located at strain ~ 0.5) is also plotted in fig. 1. It can be modelled approximately by supposing that the viscous stress there is ca. 4 times the steady state viscous stress at all shear-rates. The fit is not perfect, but then there has been no attempt to optimise it by, say, varying power-law index slightly. It suffices though to confirm that the idea that the solid phase stress reached its steady-state value at a strain ~0.5 whereas the viscous stress took much longer to settle down.

That the peak stress plotted in fig. 1 decreases with increasing shear-rate initially is telling. By contrast, Koumakis & Petekidis [6], extending the work of Pham et al. [2] on cohesive PMMA dispersions, only saw the peak stress increase with shear-rate, even though the range of Pe they explored overlapped with ours. They saw only monotonic Herschel-Bulkley type flow curves too, the two things being associated one-to-one, we think. If so, this raises the question as to why they did not see non-monotonic behaviour, especially when their stress-growth curves were not dissimilar to ours qualitatively speaking (please see fig. 3 for an example). In our case, the subsequent increase in peak stress with rate at higher rates was a viscous effect and so one possible explanation is that the overall balance of solid-phase and viscous stresses was different in their system. Their suspensions were certainly very different from our $CaCO_3$. They comprised the beautiful PMMA particles developed by ICI Paints Division [7], depletion-flocculated with polystyrene of molecular weight > $10^5$. The particle size was ca. 17 times smaller than ours, the interparticle force is estimated to have been perhaps 200 times smaller and the liquid phase of polystyrene in cis-decalin was substantially more viscous than water, hence the balance of viscous to solid-phase stress could well have been very different for one or more of several reasons.

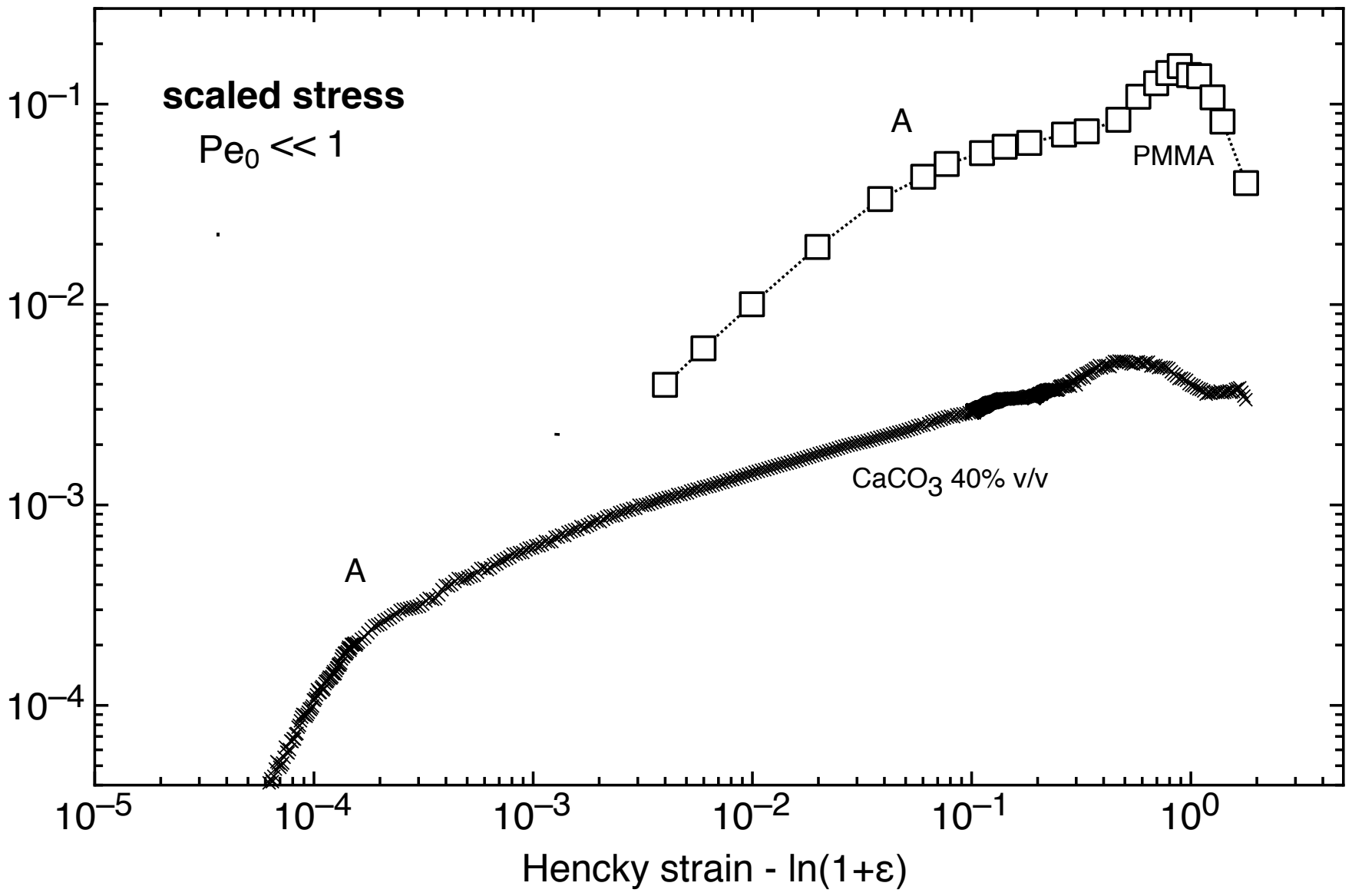

**Fig. 3** Stress growth at constant rotation rate. The nominal or "engineering" strain used to calculate the Hencky strain is given by $\varepsilon = \dot{\gamma}_N t$ where $\dot{\gamma}_N$ is the Newtonian or linear shear-rate. The stress is scaled on the shear modulus and the two curves look different qualitatively because strain-softening du to interparticle bond breakage (A) occurs at very different strains for the two systems as a result of particle size etc. PMMA data from ref. 2.

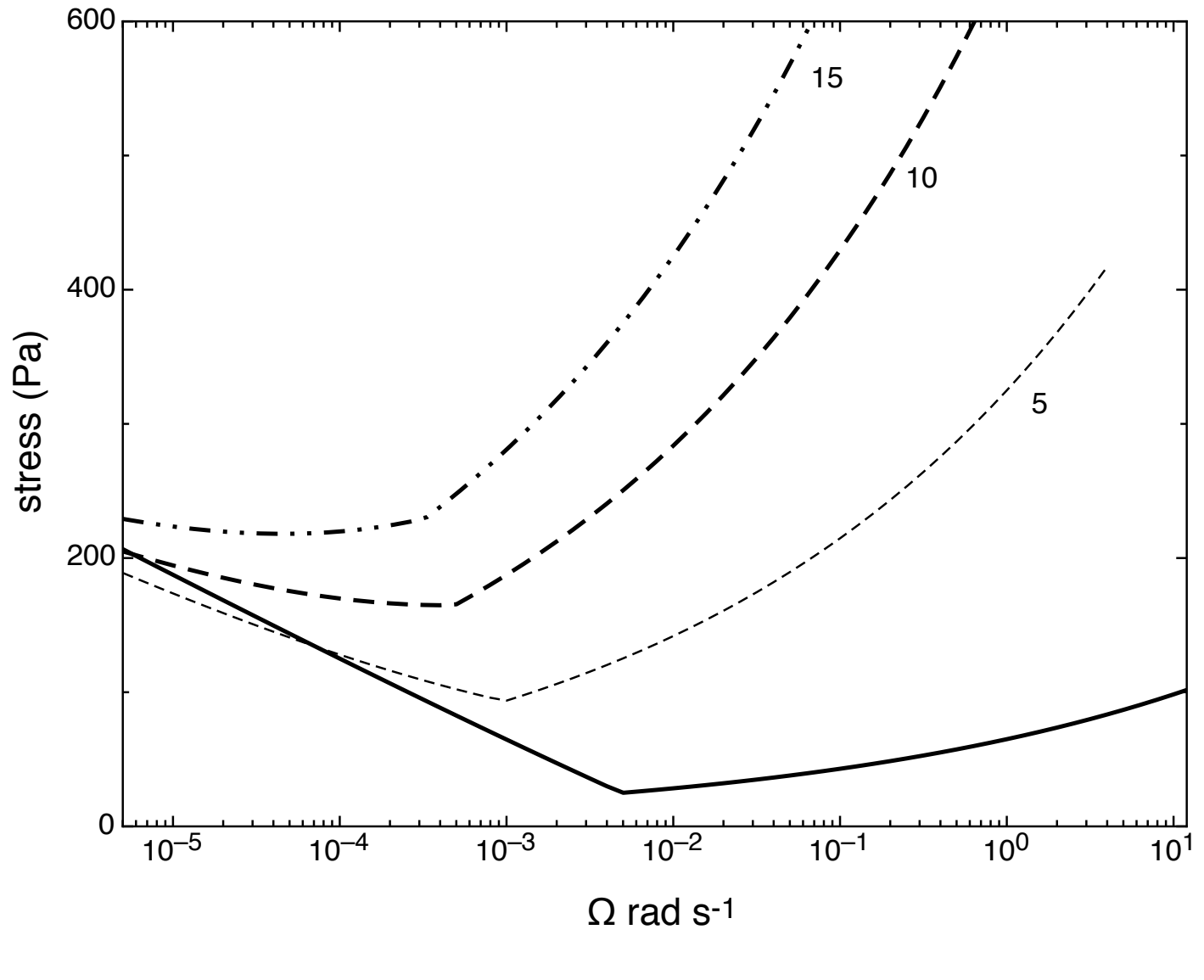

**Fig. 4** Effect of increasing the liquid phase viscosity by a factor $x$ on the fitted flow curve from fig. 1, viz. the new total stress is given by SPS($x\Omega$) + $x V_1(\Omega)$.

We do not know how to scale our flow curve data in order to compare it with theirs, especially given the very large disparities in interparticle force and particle size. Except in one respect, that is. What we can do readily is to acknowledge the difference in liquid phase viscosity: we can simply take the curve-fit of our flow curve (fig.1) and increase viscous term by a constant factor. This has been done in fig. 4. It can be seen that increases much greater than ten eliminate or disguise the strain-rate dependence of the solid phase stress. Fig. 4 suggests that the following two experiments could be of some considerable interest:

1) Replace water with corn syrup in the $CaCO_3$ system so as to increase the liquid viscosity.

2) Reduce the background viscosity of the PMMA system, either by replacing PS by, e.g. reverse-micelles, or, by replacing the solvent/PS combination by a poor solvent, thereby to induce incipient flocculation instead of depletion.

To conclude then, we think that we have hit upon a model suspension system that shows most if not all of the features of yield stress liquids observed or reported hitherto, but in one system. The large particle size (by colloidal standards) turns out to be a real advantage in two respects: It separates the characteristic “bond” and “cage-melting” strains by nearly four orders of magnitude, allowing much detail to be seen, and it renders a very wide range of Pe accessible experimentally. Furthermore, it is suspected that the strain-rate dependent yield could well be the rule rather than the exception, it being overt or not being more a matter of degree or scale, as opposed to presence or absence, and with the balance of the solid-phase and viscous stresses and their relative dependencies on Pe controlling what is seen. That is informed speculation at this stage, even if it is plausible, it is however a proposition that is readily amenable to further testing by means of range of different kinds of experiments, including those mentioned above and those others suggested in [8].

In subsequent articles we will compare the transient behaviour (cf. fig. 3) of our $CaCO_3$ system with that reported by Pham et al. [2], Koumakis & Petekidis [6] and others in more detail and we will look further at the mechanism of the deformation-rate softening. In the unlikely event that any BSR member cannot wait several months, a slide show giving more detail can be made available upon request, as can a preprints of references [3, 5 and 8] if required.

**Acknowledgements:** Funding sources are acknowledged in papers [3,5 & 8].

**References & Notes:**

# The yield stress of cohesive suspensions can be rate-dependent.

# (or, "an unlikely model system").

R. Buscall,[a,b*] H-E. Teo[a], A. D. Stickland[a], P. J. Scales[a], T. E. Kusuma[a], S. Rubasingha[a], & D. R. Lester[c,d].

[a]*Particulate Fluids Processing Centre, Dept of Chem. & Biomol. Eng., University of Melbourne, Australia 3010.*

[b] *MSACT Research & Consulting, 34 Maritime Court, Exeter, EX2 8GP, UK.*

[c] (from 1/12/14) *Dept of Chem. Eng., RMIT University, Melbourne, Australia 3001*

[d] (before 1/12/14) *CSIRO Mathematics, Informatics & Statistics, Melbourne, Australia 3190.*

The are many reasons to be interested in the constitutive rheology of colloidal suspensions, including the need to develop better engineering models of industrial processes, be they large or micro-scale. Colloidally unstable systems, i.e. aggregated or cohesive systems are of overwhelming importance in that context; stable systems being rare in technology and the environment. Of particular interest is the phenomenon of yielding, which is known to be subtle [e.g. 1-3]. In the favourable cases yield can be packaged into a lumped engineering parameter, the yield stress, although this simple parameterisation fails more generally, there being examples of systems where almost any value can be obtained, depending upon how one attempts to measure the yield stress [e.g 3,4]. One should not be too surprised perhaps, since the yield stress is no more than a convenient fiction given that stress can only be its own invariant, trivially, in one stress or strain dimension. Hence, underlying the yield stress there has always to be another criterion or set of criteria, governing the solid to liquid transition. At the simplest level one might look to stored strain energy, this being consistent with Von Mises invariant. Precisely how the true yield criterion is met in a particular test or loading protocol is then going to depend upon the viscoelasticity of the material sub-yield, and in general a combination of (CR) and controlled stress (CS) rheometry will be needed in order to develop a complete picture [e.g. 1-4].

Work on a somewhat unlikely model system will be reported, namely, a suspension of 4.5μm dia. $CaCO_3$ particles in water, coagulated at the IEP. "Unlikely", because the particle size is at the top of the colloidal range, except that this turns out to be a significant advantage in some respects. We have found this system to show just about every feature of yielding reported hitherto plus some significant new facets that have not, and probably cannot, be seen at, say, 400nm, by virtue of Peclet number (Pe).

In flow start-up experiments using step shear-rate, the stress-time curves could be scaled with shear-rate to yield curves of stress versus strain at different rates, just as could those for flocculated PMMA dispersions [5]. Transitions were seen at two characteristic strains, firstly, softening above a strain similar in magnitude to the scaled interparticle separation of the bonded particles (the "bond strain") and later a peak at a strain near unity. This type of behaviour has been reported before, notably by Petekidis et al. [3,5], working on more weakly-floccculated sub-micron PMMA particles, except that they saw the peak stress increase with shear-rate only, whereas we see decrease at first. Koumakis & Petekidis [5] have suggested that this should happen, but at very high Pe; their prediction would suggest Pe ~ $10^5$ or more for our system, whereas we see the decrease at low Pe <1. (curve A in fig.1)

The initial decrease in peak stress with shear-rate has a profound effect on the flow curves which are are highly non-monotonic (curve B in fig. 1). It also causes the suspension depicted in fig. 1, with a peak stress of ca. 200 Pa, and a "yield stress" of ~ 75 to ~ 100 Pa, depending upon how it is measured, to show an extrapolated or Herschel-Bulkley yield stress of ~zero.

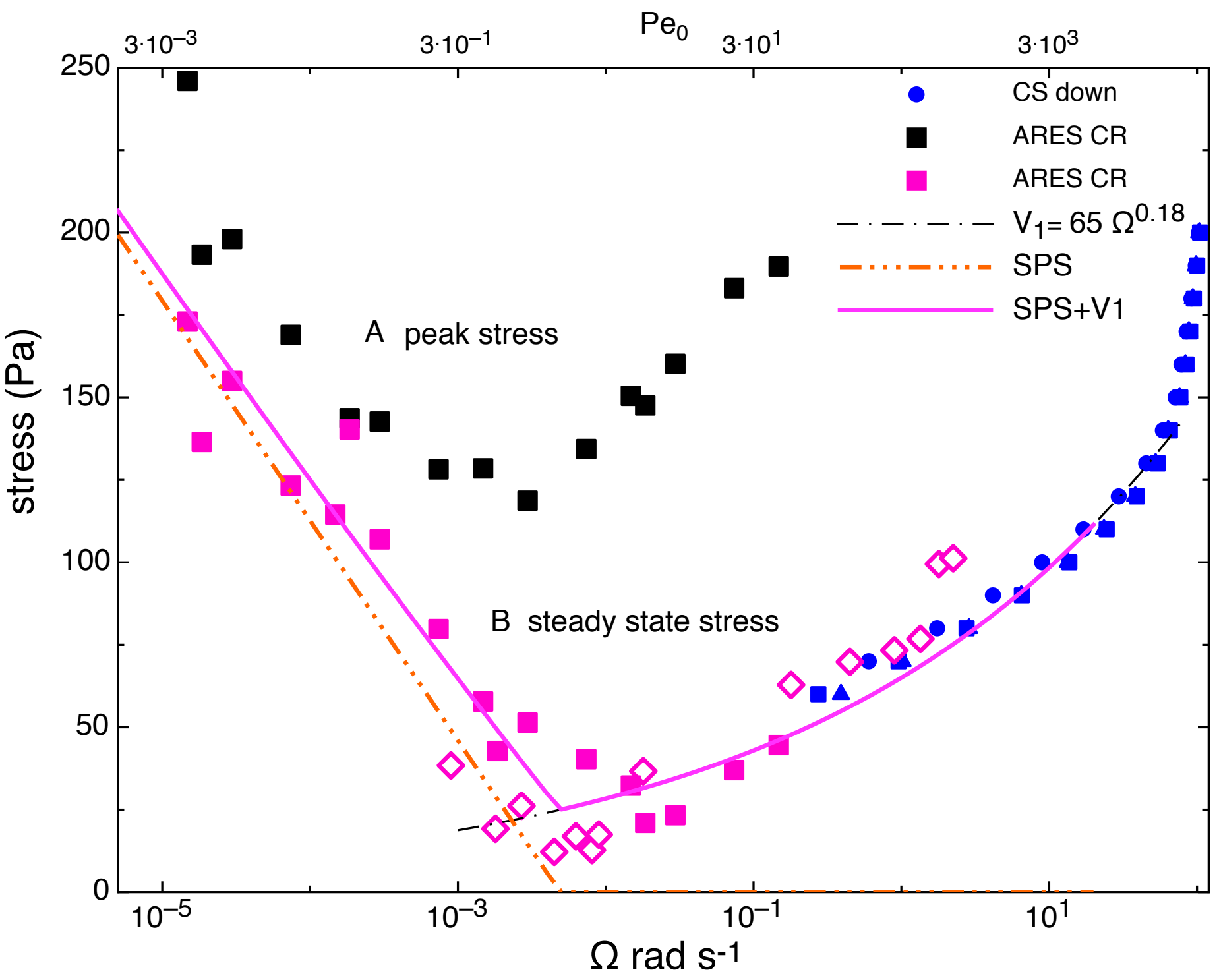

Fig. 1. A plot of stress versus angular velocity (bottom) and Pe (top) for 40%v/v Omyacarb $CaCO_3$. The steady-state curve can be decomposed by fitting into a power-law viscous part (V1) and a strain-rate thinning solid-phase part (SPS). Data taken from ref. 4.

It is helpful to separate the stress into two parts, a solid-phase part, which, were it not to be variable, one would call "the yield stress", and a viscous part. In step-strain-rate testing, the solid phase part was found to strain-soften above the bond strain, but in a Pe-dependent way. At low Pe the softening was weak and the solid stress reaches a plateau at the "cage strain" ~1 [cf. 3], whereas at higher Pe the softening became strong enough to eliminate the cage strain peak. The viscous stress was found to peak at the "cage-strain" too and this accounted for the rise in peak stress with strain-rate seen at higher strain-rates.

The strain softening extended over more than three decades in strain and it could be characterised in terms of a softening exponent (fig. 2). The exponent appears to be both Pe and volume-fraction dependent, notionally, although the latter dependence is thought to be an artefact attributable to the viscous stress growth.

One way to describe our findings in broad terms is to say that the yield strain drops from cage strain to the bond strain as Pe approaches unity. Data currently available suggest that the same thing happens when the volume-fraction is reduced below 0.25 [5,6]: the apparent yield strain, then, seems to be both strain-rate and concentration dependent.

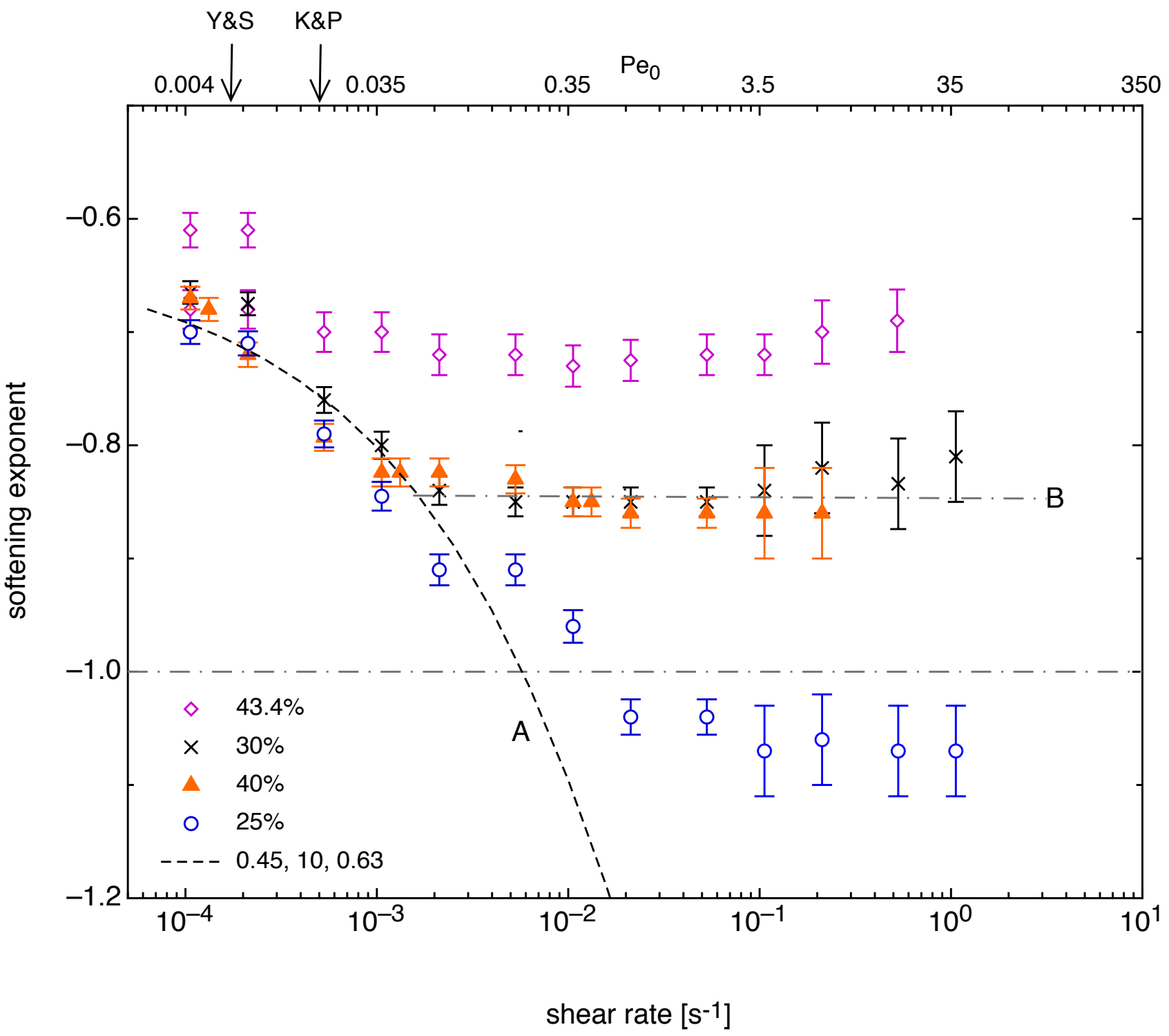

Fig. 2 Strain softening exponents for Omyacarb $CaCO_3$ at four concentrations plotted against shear rate (bottom) and Pe (top). It is thought that the true softening exponent of solid-phase stress alone follows a trend something like line A, with the apparent exponent derived from the total stress following curves more like B because of the increase in viscous stress with strain-rate.

Although Koumakis & Petekides [5] measured their step strain-rate transients over a similar Pe range, they did not see rate-dependent softening. Nor did they see non-monotonic flow curves, prompting the question "why not?". A different balance between the solid-phase and viscous stresses is one possibility. That they saw the peak stress rise with shear-rate implies that the viscous stress growth always dominated the solid-phase stress even at strains ~1. If so, then that in turn could mean perhaps that the two stress components scale very differently particle-size etc., but that remains to be seen. The liquid phase viscosity was however much higher for the PMMA systems too. The third figure shows the effect taking the fit to the flow curve in fig.1 and increasing the viscous term by a constant factor of 15, by way of illustration. The non-monotonic behaviour is then obscured, interestingly.

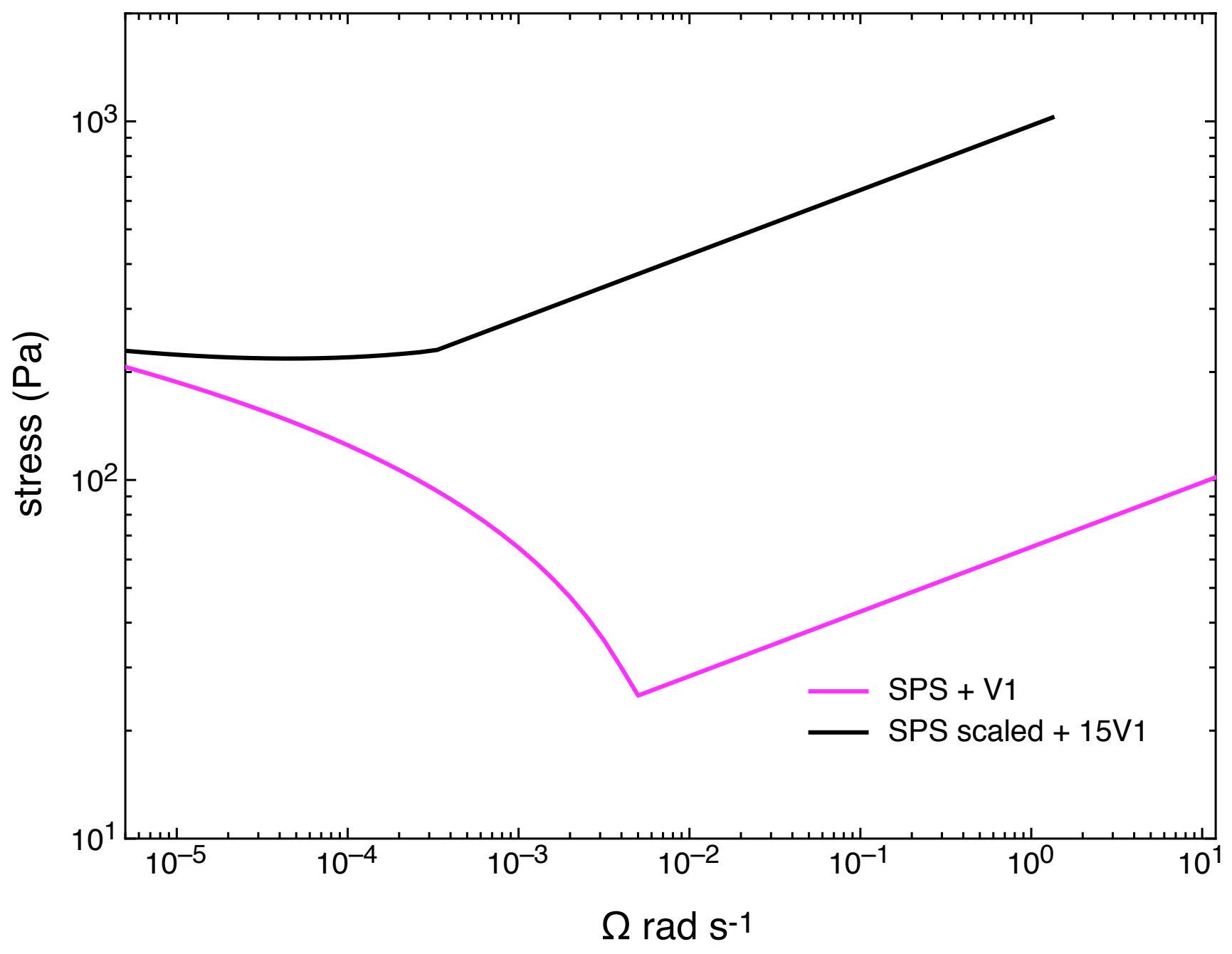

Fig. 3 shows the effect of increasing the liquid phase viscosity would have on the fit to the flow curves shown in fig.1. The viscous term has been increased by 15 and the solid-phase stress has been shifted down the rate axis to keep its dependence on Pe unaffected.

A full account of this work intended for J. non-Newtonian Fluid Mechanics is in preparation.

**Selected references:**